# Effects of Defects on Thermoelectric Properties of Carbon Nanotubes

Masato Ohnishi[1], Takuma Shiga[1], and Junichiro Shiomi[1]

[1]The University of Tokyo, Department of Mechanical Engineering
Hongo 7-3-1, Bunkyo-ku, Tokyo 113-8656, Japan

e-mail address: shiomi@photon.t.u-tokyo.ac.jp

Carbon nanotubes (CNTs) have recently attracted attention as materials for flexible thermoelectric devices. To provide theoretical guideline of how defects influence the thermoelectric performance of CNTs, we theoretically studied the effects of defects (vacancies and Stone-Wales defects) on its thermoelectric properties; thermal conductance, electrical conductance, and Seebeck coefficient. The results revealed that the defects mostly strongly suppresses the electron conductance, and deteriorates the thermoelectric performance of a CNT. By plugging in the results and the intertube-junction properties into the network model, we further show that the defects with realistic concentrations can significantly degrade the thermoelectric performance of CNT-based networks. Our findings indicate the importance of the purification of CNTs for improving CNT-based thermoelectrics.

PACS: 65.80.-g, 73.22.Pr

## I. INTRODUCTION

Over the last decades, nanoscale structures and materials have opened new possibilities to enhance thermoelectric properties. Low dimensional structures such as PbTe [1] $Bi_2Te_3$ [2], and Si/Ge superlattices [3] have been shown to give rise to high Seebeck coefficient by quantum size effect [4,5]. In addition, progress in synthesizing/fabricating nanostructured materials such as nanocrystallines [6-9], nanowires [10], and nanoporous thin films [2,11] has provided ways to reduce thermal conductivity by boundary scattering of quasi-ballistic phonons [12]. Carbon nanotubes (CNTs) are promising thermoelectric materials with both the above two merits; they are one-dimensional materials leading to the high Seebeck coefficient [13,14], and a sheet (network) consisting of CNTs is naturally a nanostructured material, where intertube junctions between CNTs reduce the thermal conductivity [15,16]. Moreover, with their flexibility, toughness, and stability, the CNT sheet can be attached to curved and movable objects [17-19] such as human body, making CNTs-based thermoelectric devices suitable for versatile applications.

While there are increasing number of reports on improvement of thermoelectric performance of sheet composed of CNTs or related carbon nanomaterials [20-23], the effect is often discussed based on a simplified picture that the thermoelectricity is mainly generated at the intertube junctions, and the body of CNT has minor contribution due to its high thermal conductivity (i.e. small temperature gradient). However, this may not be true with presence of defects that are, in practice, omnipresent in bulk CNT samples, particularly in those synthesized by using the chemical vapor deposition (CVD) methods. The crystallinity of CNTs strongly depends on CVD growth conditions; even CNT samples prepared with the same process can differ due to subtle factors that are difficult to control, such as remaining catalyst particles in a chamber [24].

While the introduction of defects to CNTs, in general, reduces both electrical and thermal conductivity (or conductance), previous works have shown that the electrical and thermal properties have different sensitivity to the type of defect; the extent of reduction varies with the defect types for electrical conductivity [25-28] but varies less for thermal conductivity [29-31]. Note that most of the works so far on the effect of crystal disorder such as defects and strain on the electronic transport properties have been done for metallic CNTs [25-28,32,33], and there are only a few works on semiconducting CNTs, which dominantly contribute to the thermoelectric performance of CNT sheets. In any case, this lack of correlation in defect-sensitivity between electrical and thermal properties complicates the effect of defects on the thermoelectric performance, since the thermoelectric figure of merit is proportional to the ratio of electrical to thermal conductivity. It is, therefore, crucial to understand the effect of defects on the thermoelectric properties to further improve the performance of CNT-based thermoelectric devices.

A few previous studies on thermoelectric performance of defective or strained carbon nanomaterials such as CNTs and graphene nanoribbons have shown that vacancies [34] or uniaxial strain [13] deteriorate the thermoelectric performance. However, in these studies, the phonon transport properties were calculated in the fully ballistic regime with atomic Green's function approach and did not consider diffusion of phonons, which makes the thermal conductance dependent on CNT length, and is known to be important for practical lengths in CNT sheets. In addition, lack in systematic analysis of the dependence on the CNT length and defect types makes it difficult to use the knowledge to estimate thermoelectric properties of CNT networks.

In this study, we systematically and comparatively study effects of defects, namely vacancies and Stone-Wales (SW) defects on thermoelectric properties of CNTs. We employ nonequilibrium molecular dynamic (NEMD) simulation to discuss the effect of the CNT length on lattice thermal properties and Green's function approach to calculate electronic transport properties. Using the knowledge, we further estimate the effect of defects on thermoelectric properties of CNT networks using a simplified model. Our calculation shows that the introduction of defects significantly deteriorates the thermoelectric performance of both individual CNTs and CNT networks because of the dominant suppression of the electron conductance.

## II. METHODOLOGY

Semiconducting (10, 0) CNTs with a relatively small diameter, $d_{cnt}$ (= 0.78 nm), are used in this study. Small-diameter semiconducting CNTs are important for the thermoelectrics because CNTs with smaller diameters have larger thermopower, the absolute value of the Seebeck coefficient [13]. In the simulated systems, a defective region is connected with right and left leads consisting of a pristine CNT as illustrated in Fig. 1. Grey fine and colored bold lines at the bottom schematics show C-C bonds around the defects before and after the structural relaxation calculation, respectively. Red colored bonds consist of atoms that clearly displaced (over 0.15 Å) around a defect due to the relaxation calculation and dashed circles indicate the region occupied by these displaced atoms. The diameters of the circle were 0.5 nm and 1.2 nm for vacancy and SW defect, respectively. The length of the defective region is varied as $L_{def}$ = 10, 50, and 100 nm. The defect concentration, $\sigma = N_{def}/N_{atom}$, is varied from 0% to 1.0%, where $N_{def}$ is the number of defects (the number of removed atoms or rotated C-C bonds for the vacancy and SW defect, respectively) and $N_{atom}$ is the number of atoms in the CNT before the introduction of defects. We introduce defects in the defective region except for both ends with the length of $0.15L_{def}$. While defects are placed randomly, their distance keeps a certain extent of distance from others (at least 1.0 nm) to avoid that regions with excessive defect concentration are generated. In the system for the NEMD calculations, the thermostated leads with the length of $L_{def}/2$ are connected with the fixed terminal layers (i.e. adiabatic boundary) consisting of a primitive unit cell. On the other hand, for Green's function calculations, the leads have a periodic and semi-infinite structure. For both calculations, relaxation calculation with the optimized Tersoff potential [35] is performed until all atomic forces become less than 0.01 eV/Å to reduce the defect induced-residual stress.

The NEMD simulations are performed using LAMMPS package [36] with the optimized Tersoff potential, which has been developed for phonon transport in carbon nanomaterials [37]. After relaxing the CNTs in canonical ensemble for more than 200 ps at 300 K, the temperature at the hot (cold) Nosé-Hoover (NH) thermostat is heated up to 310 K (cooled down to 290 K). After performing the NEMD simulation for 4 ns and the heat flow achieves a steady state, thermal properties such as heat current and temperature at each atom are obtained by averaging values at every time step for 2 ns. The heat current through the defective region is computed as $Q_{ave} = (Q_{hot} + Q_{cold})/2$, where $Q_{hot}$ and $Q_{cold}$ are the energy added to or subtracted from the hot and cold thermostats per unit time, respectively. The error between the added and subtracted energies, $|Q_{hot} - Q_{cold}|/Q_{ave}$, was up to 0.07 (less than 0.02 for most cases). The time step and damping time of thermostats are set to 0.5 fs and 50 fs for all simulations.

Lattice thermal conductivity $\kappa_{lat}$ and conductance $K_{lat}$ are then calculated as

$$\kappa_{lat} = \frac{Q_{ave}/A_{ring}}{dT/dx}, \qquad (1)$$
$$K_{lat} = \kappa_{lat} A_{ring}/L_{def},$$

where $x$ is the position along the tube axis, $dT/dx$ temperature gradient, and $A_{ring} = \pi d_{cnt} b$ the cross-sectional area of a CNT, with $b$ (= 0.34 nm) being the separation between graphite layers. To obtain the temperature gradient, we use the defective region except for both ends with the length of $0.1L_{def}$ (shaded region in Fig. 1); this means that defect is absent in the ends of the fitting region with the length of $0.05L_{def}$. This defect free region allows to prevent the generation of unusual temperature drop near the edges in the fitting region.

For the Green's function calculations, we use a tight-binding method [33], where the hopping integral for π-orbital between carbon atoms is attenuated exponentially with increasing the bond length [38]. Green's function and the transmission function of the defected region are obtained as

$$G_{def} = [(E + i\eta) - H_{def} - \Sigma_L - \Sigma_R]^{-1}, \qquad (2)$$
$$\Theta(E) = \text{tr}\left[\Gamma_L G_{def} \Gamma_R G_{def}^{\dagger}\right],$$

where $E$ is the energy of incident electron to the defective region, $\eta$ is the infinitesimal, $H_{def}$ is Hamiltonian matrix of the defective region, $\Sigma_{L(R)}$ is the self-energy matrix of the left (right) lead, and $\Gamma_{L(R)} = i\left[\Sigma_{L(R)} - \Sigma_{L(R)}^{\dagger}\right]$. The electric current $I(V)$ and electronic thermal current $J(V)$ under the bias voltage $V$ through the defective region are obtained using Landauer-Büttiker formula [39]:

$$I(V) = \frac{2e}{h} \int dE\, \Theta(E) \left[f_L(E - \mu_L) - f_R(E - \mu_R)\right],$$
$$J(V) = \frac{2e}{h} \int dE\, \Theta(E) \left[f_L(E - \mu_L) - f_R(E - \mu_R)\right](E - \mu), \tag{3}$$

where $e$ is the electron charge, $h$ is the Plank constant, $f_{L(R)}$ is the Fermi-Dirac distribution function of the left (right) lead, which is also a function of temperature, and $\mu$ is the chemical potential, which can be tuned with gating or doping. The temperature is set to 300 K in all the simulations. Under the linear response approximation, i.e. when the differences of the chemical potential, $\Delta\mu = \mu_L - \mu_R$, and the temperature difference $\Delta T$ between both leads are infinitesimally small, we can obtain thermoelectric properties as follows [13,16].

The electronic conductance:

$$G_{el} = -\left.\frac{I}{\Delta V}\right|_{\Delta T=0} = e^2 A_0. \tag{4}$$

The Seebeck coefficient:

$$S = -\left.\frac{\Delta V}{\Delta T}\right|_{I=0} = \frac{A_1}{eTA_0}. \tag{5}$$

The electronic thermal conductance:

$$K_{el} = \left.\frac{J}{\Delta T}\right|_{I=0} = \frac{A_0 A_2 - A_1^2}{TA_0}. \tag{6}$$

Here, $A_n$ is defined as the following integral,

$$A_n = \frac{2}{h} \int dE\, \Theta(E) \left(-\left.\frac{\partial f}{\partial E}\right|_{E=\mu}\right)(E-\mu)^n. \tag{7}$$

Using Eqs. (1) and (4)-(6), we can obtain the power factor, $P = S^2 G_{el}$, and the thermoelectric figure of merit, $Z_{cnt}T = S^2 G_{el}/(K_{lat} + K_{el})$. For each defect concentration $\sigma$, electronic calculations are performed for five different random defect configurations, and the obtained electron properties (Eqs. (4)-(6)) are averaged. Note that, in case of thermal transport calculations, NEMD simulations were performed for single configuration for each defect concentration because the sensitivity to the defect configuration is much smaller than for electrical transport properties. This was checked by analyzing five configurations for some cases of $\sigma$ in 10 nm-CNTs, and the resulting variations in $\kappa_{lat}$ were sufficiently small (20% at most and less than 10% in most cases). This error will further diminish by fitting $\kappa_{lat}$ as a function of $\sigma$.

The fundamental difference between the methods of thermal and electrical calculations is worth mentioning. In the Green's function method, the electron-phonon and electron-electron scattering are neglected because the electron mean free path (MFP) of pristine semiconducting CNTs is known to reach 200 nm at room temperature [40], which is sufficiently longer than the CNTs used in this study. The MFP is expected to be shorter due to electron-phonon or -electron scattering induced by the localized phonons and electrons around the defects but our calculations should be valid at least for low defect concentrations. On the other hand, in the NEMD simulation, it is important that the anharmonic phonon-phonon scatterings are considered because contribution to thermal conductivity comes from phonons with a wide range of frequencies including those with MFPs shorter than the CNTs used in this study [41].

### III. LATTICE THERMAL TRANSPORT PROPERTIES

#### A. Thermal conductivity of pristine CNTs

First we validate $\kappa_{lat}$ obtained in this study. While $\kappa_{lat}$ of pristine (10, 0) CNTs obtained in this study are 170, 530, and 750 W/m-K for $L_{def}$ = 10, 50, and 100 nm, respectively, it is known that $\kappa_{lat}$ of CNTs varies widely depending on different factors as follows. Salaway *et al.* show that the optimized Tersoff potential estimates $\kappa_{lat}$ of CNTs larger than other empirical interatomic potentials [37]. In fact, $\kappa_{lat}$ of 100 nm-(10, 10) CNT (950 W/m-K), which is additionally calculated for this study, is larger than

$\kappa_{\text{lat}}$ calculated with the adaptive intermolecular reactive empirical bond order (AIREBO) potential (200 W/m-K) [37], Brenner potential (210 W/m-K) [42], and simplified Brenner potential (330 W/m-K) [37,41]. While the choice of interatomic potential is still controversial, the optimized Tersoff potential is used in this study because this potential reproduces experimentally observed phonon properties of CNTs or graphite more accurately such as phonon dispersion, group velocities [43,44], and thermal conductivity [45].

Conditions of NH thermostats (length, temperature difference, and damping time) also affect the magnitude of $\kappa_{\text{lat}}$ [41]. $\kappa_{\text{lat}}$ of CNTs increases with the length of thermostats because of the increase in the number of phonons generated in thermostats with their length and converges when the length of thermostats approaches half of $L_{\text{def}}$ [41]. $\kappa_{\text{lat}}$ of CNTs calculated in this study (750 W/m-K for 100 nm-(10, 0) CNT), therefore, are larger than those obtained by Cao et al. (400 W/m-K for 100 nm-(10, 0) CNT) using shorter thermostats (2 nm to 10 nm) [46]. In fact, when changing the length of thermostat from 50 nm to 2 nm, we observed 30% reduction of $\kappa_{\text{lat}}$ for 100 nm-(10, 0) CNT (from 750 W/m-K to 520 W/m-K). In contrast, changing the damping time and the temperature difference between thermostats ($T_{\text{hot}} - T_{\text{cold}}$) from (50 fs, 20 K) to (1 ps, 60 K), the same conditions as those used by Sevik et al. [47], did not affect $\kappa_{\text{lat}}$ of CNTs. Here, $\kappa_{\text{lat}}$ obtained by Sevik et al. are almost half of $\kappa_{\text{lat}}$ obtained in this study and are comparable with $\kappa_{\text{lat}}$ obtained by Salaway et al. and Cao et al. while the former used longer thermostats (constant at 50 nm) than those used by the latter. The cause of the discrepancy is not clear at this point, and may be due to differences in more detail methodology of the simulations that cannot be judged from the information available in the paper, however, further investigation of the discrepancy is beyond the scope of this paper. Nevertheless, the effect of defects on $\kappa_{\text{lat}}$ of CNTs is consistent with that observed by Sevik et al. as will be discussed below.

### B. Thermal conductivity of defective CNTs

Figure 2 shows the change in $\kappa_{\text{lat}}$ of (10, 0) CNTs with different $L_{\text{def}}$, 10 nm (black circle), 50 nm (blue square), and 100 nm (orange triangle), due to the introduction of (a) vacancies and (b) SW defects. Here, $\kappa_{\text{lat}}(\sigma)$ can be written as $\kappa_{\text{lat}}(\sigma) = c_{\text{ph}} v_g \Lambda_{\text{tot}}(\sigma)$ with $c_{\text{ph}}$ being heat capacity, $v_g$ phonon group velocity, and $\Lambda_{\text{tot}}$ phonon MFP in defective CNTs. $\Lambda_{\text{tot}}$ satisfies the Matthissen's rule: $\Lambda_{\text{tot}}^{-1} = \Lambda_{\text{prist}}^{-1} + \Lambda_{\text{def}}^{-1}$, where $\Lambda_{\text{prist}}$ and $\Lambda_{\text{def}}$ are the MFPs in the pristine CNT and induced by defects. Assuming that $c_{\text{ph}}$ and $v_g$ are independent on $\sigma$ and $\Lambda_{\text{def}}$ is proportional to $\sigma^{-\beta}$, we can obtain the fitting equation, $\kappa_{\text{lat}}(\sigma) = \kappa_{\text{lat}}(0)\left[1 + \alpha \kappa_{\text{lat}}(0)\sigma^{\beta}\right]^{-1}$ [31], where $\alpha$ and $\beta$ are fitting parameters. Solid (broken) lines in Fig. 2 show fitting curves for vacancy (SW defect) with this relationship. In Fig. 2(a), fitting curves for SW defect (broken line) are also shown to compare with data for vacancy and its inset shows a blow-up of region at high $\sigma$.

The introduction of defects significantly decreases $\kappa_{\text{lat}}$ particularly at low $\sigma$ (< 0.2%) for both defects as also shown in previous studies [31,47]; $\kappa_{\text{lat}}$ is reduced by half at $\sigma$ = 0.14% (0.11%), 0.062% (0.092%), and 0.060% (0.078%) for 10, 50, and 100 nm-CNTs with vacancies (SW defects), respectively. Vacancies decrease $\kappa_{\text{lat}}$ of CNTs more effectively than SW defects [29,30] because of the absence of C-C bonds around vacancies, which obviously diminish short-wavelength phonons directly, while their difference is not obvious in short CNTs as shown in Fig. 2(a). Figure 2(a) also shows that, for SW defect, the dependence of $\kappa_{\text{lat}}$ on $L_{\text{def}}$, one of the ballistic features of phonon transport [48], remains even at high $\sigma$ (≈ 1.0%) for CNTs with $L_{\text{def}} \leq 100$ nm. Sevik et al. also shows that for longer CNTs (200-600 nm) the length dependence remains to some extend for SW defects of $\sigma$ = 0.6% [47]. On the other hand, the $L_{\text{def}}$-dependence of $\kappa_{\text{lat}}$ diminishes more rapidly for vacancy: at $\sigma$ < 0.3% for $L_{\text{def}}$ = 200-600 nm (Ref. [47]), $\sigma$ ≈ 0.8% for $L_{\text{def}}$ = 50-100 nm, and $\sigma$ > 1.0% for $L_{\text{def}}$ < 50 nm. This result can be understood from the analysis with atomic Green's function method by Sevik et al. [47]. Their analysis shows that phonon MFPs due to vacancy-induced elastic scattering decrease from 100 nm to 20 nm when $\sigma$ increases from 0.1% to 1.0% for most phonons (phonons with frequency above 400 cm$^{-1}$) while MFPs of lower frequency (< 200 cm$^{-1}$) phonons exceed 1 $\mu$m even under high $\sigma$ (≈ 1.0%). Therefore, while the $L_{\text{def}}$-dependence of $\kappa_{\text{lat}}$ diminishes at low $\sigma$ (≈ 0.1%) of vacancies when 100 nm ≤ $L_{\text{def}}$ ≤ 1 $\mu$m, it remains even at high $\sigma$ (> 0.8%) of vacancies for shorter CNTs ($L_{\text{def}}$ < 100 nm) because of the comparable length of MFPs with CNT length. The reason why the ballistic feature is more observable in shorter CNTs (Fig. 2) than in longer CNTs (Ref. [47]) should be the same as the above discussion on vacancy.

### IV. ELECTRON TRANSPORT PROPERTIES

### A. Transmission function

Since electron contribution to thermoelectric properties is determined by $\Theta(E)$ as shown in Eqs. (2)-(7), the change in $\Theta(E)$ due to defects are discussed here. Figure 3 shows $\Theta(E)$ of 100-nm CNTs for (a) vacancies and (b) SW defects with $\sigma$ varying from 0.0% (blue line) to 0.1% (red line) with the equal interval of $\sigma$ (corresponding to $N_{\text{def}}$ = 0 to 9). The Fermi level is set to 0 eV. Broken lines indicate the peak chemical potentials, $\mu$ giving the maximum $P$ (see Fig. 4(d)), for p-/n-type pristine CNT, defined as $\mu_{\text{p/n},0} = -/+0.38$ eV. $\mu_{\text{p/n},0}$ is located at potential levels slightly higher/lower than the valence/conduction band edge ($E = -/+0.41$ eV) [34]. Blow-ups on the bottom show $\Theta(E)$ at the marked region around $\mu_{\text{p/n},0}$. While both of vacancies and SW defects suppress $\Theta(E)$ significantly, their effects on $\Theta(E)$ are different in some aspects. Vacancies selectively suppress $\Theta(E)$ at band edges, corresponding to the energy levels of Van-Hove singularities. This selective suppression of $\Theta(E)$ due to vacancy can be attributed to the generation of quasi-bound states [25,49], which are generated mostly at energy levels near Van-Hove singularities (band edges) and suppress $\Theta(E)$ at the corresponding energy levels. On the other hand, SW defects suppress $\Theta(E)$ in the overall energy range somewhat keeping the original step-like feature. Here, in our additional calculation, we observed that bond distortions without adding any defect decrease $\Theta(E)$ at overall energy levels rather than at specific energy levels. This indicates that the suppression of $\Theta(E)$ in the overall energy range due to SW defects can be described by broad areas of bond distortions as illustrated in Fig. 1. These differences between effects of vacancy and SW defect on $\Theta(E)$ mainly affect the change in $S$ due to defects as shown below.

## B. Fluctuation of electron contributions to thermoelectric properties

Figure 4 shows different thermoelectric properties, (a) $S$, (b) $G_{el}$, (c) $K_{el}$, and (d) $P$ of CNTs with vacancies (top) and SW defects (bottom). The range of $\sigma$ and its color notation are the same as those in Fig. 3. Insets in Figs. 4(a), 4(b), and 4(d) are the blow-ups around $\mu_{p/n,0}$, denoted by broken lines while insets in Fig. 4(c) shows $K_{el}/(G_{el}TL)$ with $L$ being the Lorentz number. The introduction of vacancies increases $|S|$ around $\mu_{p/n,0}$ while SW defects do not substantially affect $|S|$ as shown in Fig. 4(a) (this trend can be seen clearer in Fig. 6(a)). The trend can be attributed to the aforementioned change in $\Theta(E)$ [34]; while $\partial f/\partial E$, a window function in the denominator of Eq. (5), takes its peak at $E = \mu$, $(\partial f/\partial E)(E-\mu)$, a window function in the numerator of Eq. (5), takes at $E \neq \mu$ ($E = \mu \pm 0.04$ eV for 300 K). In addition, because the former attenuates more rapidly than the latter with increasing $|E-\mu|$, the denominator of Eq. (5) is dominated by $\Theta(E)$ around $E = \mu$ compared with the numerator. Therefore, considering $S$ at $\mu_{p/n,0}$, vacancies, which selectively suppress $\Theta(E)$ near the band edge (near $E = \mu_{p/n,0}$), can mainly reduce the denominator of Eq. (5) and, thus, increase $|S|$. On the other hand, since SW defects decrease $\Theta(E)$ in the overall energy range and decreases both the denominator and the numerator, effects of their changes on $S$ are canceled out and, thus, $S$ does not change substantially. The above discussion allows the further expectation that electron disorders can increase $S$ when the disorders do not cause bond distortion but affect electronic states mainly at the band edge (e.g. adatoms).

While $S$ varies with a relatively complex manner, $G_{el}$ and $K_{el}$ simply follow the change in $\Theta(E)$ as shown in Figs. 4(b) and 4(c); $G_{el}$ and $K_{el}$ decrease with $\Theta(E)$ with increasing $\sigma$ for both vacancy and SW defect. Compared with the increase in $S$, the reduction of $G_{el}$ and $K_{el}$ are more significant (also see Fig. 6). This result shows that $S$ is dominated by electronic structures at lead regions while $G_{el}$ and $K_{el}$ by electron scattering at the defective region. $G_{el}$ and $K_{el}$ near $\mu_{p/n,0}$, the energy range dominating the thermoelectric properties, decrease significantly with increasing $\sigma$ (inset in Fig. 4(b) for $G_{el}$). As for $K_{el}$, $K_{el}$ follows the Wiedemann-Franz law except for around energy levels corresponding to the Van-Hove singularities as well as the band gap. The electron contribution to the heat transport, $K_{el}$, is much less than the lattice contribution, $K_{lat}$, regardless of $\sigma$; for 100 nm-CNTs, $K_{el}(\mu_{p,0})/K_{lat}$ was 0.09 and 0.01 for $\sigma = 0\%$ and 0.1% of vacancy, respectively.

With increasing defects, $P$, determined by $S$ and $G_{el}$, finally reduces significantly as shown in Fig. 4(d) for all cases including the case of vacancy, which enhances $|S|$. This result shows that the enhancement of $|S|$ is overwhelmed by the large reduction of $G_{el}$, despite that $P$ is quadratic to $S$ (linear to $G_{el}$) (this trend will be discussed again in Fig. 6). Here, one can notice that, with decreasing $P$, the peak (optimized) chemical potential, $\mu_{p/n,opt}$, that gives the maximum $P$ for each defective CNT may shift from $\mu_{p/n,0}$. While it should be appropriate to adopt $\mu_{p/n,0}$ as the representative $\mu$ for the pristine CNT, there are two alternatives for defective CNTs: one is to adopt the same $\mu_{p/n,0}$ assuming that $\mu$ remains the same in the process of introducing defects, and the other is to take $\mu_{p/n,opt}$ for each defective CNT reflecting the maximum possible $P$. Since neither of the representative $\mu$ is universally appropriate and the defects similarly affect p- and n-type CNTs properties, in the followings, we mainly focus on p-type CNTs and evaluate the properties for both $\mu_{p,0}$ and $\mu_{p,opt}$ that will be simply denoted by $\mu_{opt}$ and $\mu_0$ hereafter.

$\mu_{opt}$ indeed changes with increasing $\sigma$ particularly for vacancy as shown in Fig. 5(a). For vacancy in 100 nm-CNT, $\mu_{opt}$ decreases from $\mu_0$ (–0.38 eV) and saturates to $\mu = -0.54$ eV at $\sigma \approx 0.5\%$. The magnitude of $\sigma$ at which $\mu_{opt}$ saturates increases with increasing $L_{def}$. This reflects the fact that the same magnitude of $\sigma$ decreases $\Theta(E)$ more effectively in longer CNTs in the ballistic regime because increasing length for constant $\sigma$ means larger $N_{def}$. The change in $\mu_{opt}$ can be understood from Fig. 5(b) showing different thermoelectric properties of 100 nm-CNT with $\sigma = 0.00$, 0.02, and 0.04% ($N_{def} = 0$, 3, and 6). In Fig. 5(b), the units for $|S|$, $G_{el}$, and $P$ are normalized as V/(5000K), S/5000, and pW/K$^2$, respectively. This figure summarizes aforementioned trends; with decreasing $\Theta(E)$ due to the introduction of vacancies, $S$ increases (this trend is not clear in this figure because of its slight increase) and $G_{el}$ and $P$ decrease around $\mu_0$. Because of the competing effect of vacancy on $S$ and $G_{el}$ around $\mu_0$, the introduction of vacancies causes the shift of $\mu_{opt}$ toward high-doping level as well as the decrease in the magnitude of $G_{el}$ and $P$.

The changes in $|S|$, $G_{el}$, and $P$ with optimizing $\mu$ in terms of $P$ are shown in Fig. 6 to clarify impacts of defects on $S$ and $G_{el}$. Solid and broken lines represent data at $\mu_{opt}$ and $\mu_0$, respectively. While $|S|$ increases a few times for vacancy at $\mu_0$ (up to 4 times for 100-nm CNT), $P$ decreases with increasing $\sigma$ (more than four orders of magnitude) because of orders of reduction of $G_{el}$ (up to five orders of magnitude for 100-nm CNT). While without the optimization of $\mu$ the effect of vacancy on thermoelectric properties is larger than that of SW defect, the $\mu$-optimization can recover $G_{el}$ as well as $P$ of CNTs with vacancies up to the same orders or even higher (in the case of 100 nm-CNT) than those of CNTs with SW defect. Here, it is interesting to note that after the optimization of $\mu$, $|S|$ at $\mu_{opt}$ of CNTs with vacancies remains almost constant ($\approx 0.2$ mV/K) regardless of $\sigma$. In the case of SW defect (right column of Fig. 6), while the introduction of the defects does not change $|S|$, the significant reduction of $G_{el}$ due to defects (up to five order of magnitude for 100-nm CNT) deteriorates $P$ by the same orders as $G_{el}$. As a result, our findings reveal that the deterioration of $P$ due to the introduction of defects is strongly dominated by the orders of suppression of $G_{el}$ following that of $\Theta(E)$ although $|S|$ increases in the case of vacancy.

### C. Figure of merit of individual CNTs

Finally, combining thermal and electron transport properties calculated in the above, we calculated $ZT$ of individual CNTs. Figure 7 shows $Z_{cnt}T$ of CNTs with (a) vacancies and (b) SW defects of the length of 10 nm (black circle), 50 nm (blue triangle), and 100 nm (orange diamond) at $\mu_0$ (dashed line) and $\mu_{opt}$ (solid line). Insets show the change in $Z_{cnt}T$ in low $\sigma$ region with linear scale. For the pristine CNTs ($\sigma = 0\%$), $Z_{cnt}T$ increases with the CNT length (0.06 for 10 nm, 0.08 for 50 nm, and 0.1 for 100 nm). This is because, for the pristine CNTs, while the electronic transport properties do not depend on the CNT length in the fully ballistic regime, the phonon transport, whose anharmonicity is not negligible even in short CNTs, degrades with increasing the CNT length. However, because electronic transport properties, particularly $G_{el}$, of longer CNTs are fluctuated more sensitively due to defects, $Z_{cnt}T$ of longer CNT reduces more significantly and the length dependence of $Z_{cnt}T$ reverses at $\sigma \approx 0.02\%$; i.e. $Z_{cnt}T$ decreases with increasing $L_{def}$ under $\sigma$ exceeding 0.02%. As a result, while both of $K_{lat}$ and $P$ reduce due to defects, the change in $Z_{cnt}T$ due to defects is dominated by electronic properties, particularly $G_{el}$, which reduces orders of magnitude with increasing $\sigma$.

### V. THERMOELECTRIC PROPERTIES OF CNT-BASED NETWORKS

Using calculated results in the above, we estimate thermoelectric properties of sheets composed of CNTs with vacancies. Thermal conductivity of three-dimensional networks composed of randomly-dispersed straight CNTs can be calculated as [50,51]

$$\kappa_{net} = \frac{\kappa_{net}^\infty}{1 + L_{def}N_J/(12L_{eff}^{th})} = \frac{\kappa_{net}^\infty}{1 + F(L_{def})/L_{eff}^{th}}, \quad (8)$$

where $F(L_{def}) = L_{def}N_J/12$ is a function increasing monotonically with $L_{def}$ and $L_{eff}^{th} = (K_{lat} + K_{el})/K_{cc}$ with $K_{cc}$ being thermal conductance at intertube junctions is the effective length of thermal conductance at intertube junctions. $\kappa_{net}^\infty$ and $N_J$ are the thermal conductivity of networks composed of CNTs with infinite thermal conductivity and the mean number of junctions per CNT, respectively:

$$\kappa_{net}^\infty = \frac{K_{cc}}{d_{cnt}} \frac{\pi \bar{n}_V^2}{18} \left(1 + 8\bar{d}_{cnt} + 20\bar{d}_{cnt}^2 + 24\bar{d}_{cnt}^3 + 9.6\bar{d}_{cnt}^4\right), \quad (9)$$

$$N_J = \pi \bar{n}_V \left(1 + 4\bar{d}_{cnt} + \frac{8}{3}\bar{d}_{cnt}^2\right), \quad (10)$$

where $\bar{d}_{cnt} = d_{cnt}/L_{def}$, $\bar{n}_V = n_V L_{def}^2 d_{cnt}/2$, and $n_V$ is the volume number density of CNTs. The volume density is fixed at 20%, corresponding to 0.35 g/cm$^3$. We used the same formula and abbreviations (e.g. $L_{eff}^{el} = G_{el}/G_{cc}$ with $G_{cc}$ being electron conductance at intetutbe junctions) to discuss electrical properties of CNT networks. $K_{cc}$ (= 50 pW/K) is calculated with the empirical formula based on NEMD simulations [52] (see Appendix) and experimentally observed $G_{cc}$ (= 3.8 μS) with small-diameter SWNTs ($d_{cnt} < 3$ nm) [53] is employed to obtain the electrical conductivity of networks, $\lambda_{net}$. As for Seebeck coefficient, because Seebeck coefficient of CNT networks, $S_{net}$, is not sensitive to the network condition (e.g. morphology and number of contacts) and is dominated by $S$ of individual CNTs [23], we use $S$ of individual CNTs as $S_{net}$ for simplicity and obtain the figure of merit of networks as $Z_{net}T = S^2 \lambda_{net}/\kappa_{net}$. For electron properties ($G_{el}$ and $S$), values at $\mu_{opt}$ are used in this estimation.

Calculated thermoelectric properties of networks composed of CNTs with the length of 10 (circle), 50 (square), and 100 (triangle) nm are shown in Fig. 8. Because the distance between defects ranges from ≈ 20 nm (for CNTs with relatively high $\sigma$ [54]) to sub-microns (for highly-purified CNTs fabricated with a CVD method under high temperature condition [55,56]), we calculate values in the plausible range of $\sigma$, in which the effective averaged distance between defects, $L_{def}/N_{def}$, exceeds ≈ 10 nm. Figure 8 shows $ZT$ values of CNT networks increase compared with those of individual CNTs (Fig. 7). Here, $Z_{net}T$ can be written as

$$Z_{net}T = \frac{L_{eff}^{th}}{L_{eff}^{el}} \frac{1 + F(L_{def})/L_{eff}^{el}}{1 + F(L_{def})/L_{eff}^{th}} Z_{cnt}T \quad (11)$$

This equation shows that $L_{eff}^{th} > L_{eff}^{el}$ results in $Z_{net}T > Z_{cnt}T$ as shown in Figs. 7 and 8 ($L_{eff}^{th}/L_{eff}^{el} = 14$ for 100-nm pristine CNT in the present simulation). Equation (11) also shows that, under the same density of CNTs, when $L_{eff}^{th} > L_{eff}^{el}$ and both do not depend on $L_{\text{def}}$, networks of shorter CNTs show better thermoelectric performance because $d(Z_{\text{net}}T)/dL_{\text{def}} < 0$. This trend can be observed in Fig. 8; $Z_{\text{net}}T$ for shorter CNTs is larger than $Z_{\text{net}}T$ for longer CNTs in whole range of $\sigma$. When the defects are introduced, the $Z_{\text{net}}T$ follows the strong suppression of $Z_{\text{cnt}}T$ shown in Fig. 7 while the change in other terms are complicated; i.e. the first and second fractional terms of Eq. (11) increases and decreases, respectively, because defects decrease $G_{\text{el}}$ more effectively than $K_{\text{lat}}$ ($L_{\text{eff}}^{\text{el}}$ decreases stronger than $L_{\text{eff}}^{\text{th}}$ with increasing $\sigma$). Consequently, the introduction of defects decreases $Z_{\text{net}}T$ by ≈ 40% (e.g. 0.78 for $L_{\text{def}}/N_{\text{def}}$ = 0 nm to 0.50 for $L_{\text{def}}/N_{\text{def}}$ = 20 nm for 100-nm CNTs). Our estimation indicates that the purification of CNTs and the usage of shorter CNTs can increase $Z_{\text{net}}T$. Furthermore, considering that the effects of defects increase with the CNT length, the deterioration of the thermoelectric performance is more crucial in CNT-based networks with longer CNTs.

## VI. CONCLUSIONS

We theoretically investigated effects of defects, vacancies and SW defects, on the thermoelectric properties of semiconducting CNTs and CNT-based networks. We found that vacancies can increase in the Seebeck coefficient of individual CNTs (by up to four times) due to the selective suppression of the transmission function at energy levels corresponding to Van Hove singularities. However, significant suppression of electron conductance regardless of the type of defect (by up to five orders) overwhelms the increase in Seebeck coefficient. As for the comparison of effects of the defects, while the reduction of $ZT$ due to vacancies is larger than that due to SW defects at the fixed chemical potential, since the $\mu$-optimization functions more efficiently for vacancy, $ZT$ for vacancy is larger than $ZT$ for SW defect with the $\mu$-optimization. Further calculations on effects of defects on thermoelectric performance of CNT-based networks show that the purification of CNTs and the usage of shorter CNTs can effectively increase $ZT$ of CNT-networks. Our findings show concrete ways to enhance the performance of CNT-based thermoelectric devices.

## ACKNOWLEDGEMENT

This study is partially supported by Thermal Management Materials and Technology Research Association (TherMAT) and JSPS KAKENHI Grant Number JP26709009. The numerical calculations in this work were carried out on the facilities of the Supercomputer Center, Institute for Solid State Physics, The University of Tokyo and the TSUBAME2.5 supercomputer in the Tokyo Institute of Technology.

## APPENDIX: THERMAL CONDUCTANCE AT INTERTUBE JUNCTIONS

As derived in Ref. [52], thermal conductance at intertube junctions can be calculated by using the empirical formula.

$$K_{\text{cc}} = (AM_{\text{eff}}^B + C)N_{\text{eff}}, \quad (12)$$

where A = –1.62 × 10⁻¹¹ pW/K, B = 10.86, and C = 0.2154 pW/K. $N_{\text{eff}}$ and $M_{\text{eff}}$ are total effective number of interatomic intertube interactions and effective interatomic intertube interactions per atom in the contact region, respectively. The contribution from a pair of $i$th atom in a tube and $j$th atom in another tube is derived by the following Lennard-Jones potential form equation:

$$n(r_{ij}) = \begin{cases} 1, & r_{ij} < r_m \\ 2(r_m/r_{ij})^6 - (r_m/r_{ij})^{12}, & r_m \leq r_{ij} \leq r_c \\ 0, & r_c < r_{ij} \end{cases} \quad (13)$$

where $r_{ij}$ is the distance between $i$th and $j$th atoms, $r_m = 2^{1/6}\sigma_{\text{int}}$ is the distance corresponding to the minimum of the potential, $\sigma_{\text{int}}$ = 3.4 Å is the length parameter of the Lennard-Jones potential, and $r_c$ = 10 Å is the cutoff length.

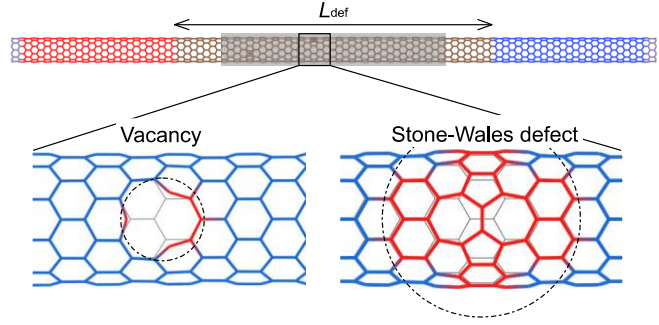

FIG. 1. (Color online) Schematic of a CNT with defects: vacancy and SW defect. The defective region connects with pristine CNT leads at the both ends. For the NEMD simulation, the length of the leads is a half of that of the defective region ($L_{def}/2$) and are terminated by fixed layers. For Green's function method, on the other hand, the leads are defined to be semi-infinite and periodic. Dashed circles in the bottom panels show the region in which atoms are displaced over 0.15 nm due to the introduction of defects.

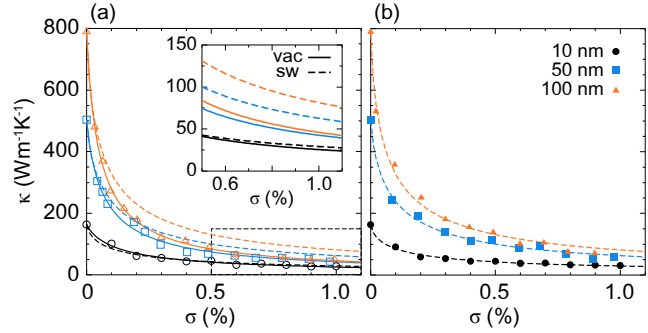

FIG. 2 (Color online) Change in thermal conductivity of (10, 0) CNTs due to (a) vacancy and (b) SW defect. The symbols, black circle, blue square, and orange triangle represent different CNT lengths, 10, 50, and 100 nm, respectively. Solid line in (a) shows the fitting line for vacancy and broken line in (a) and (b) for SW defect. The inset in (a) shows a blow-up at high defect concentration ($0.5\% \leq \sigma \leq 1.0\%$), the marked region.

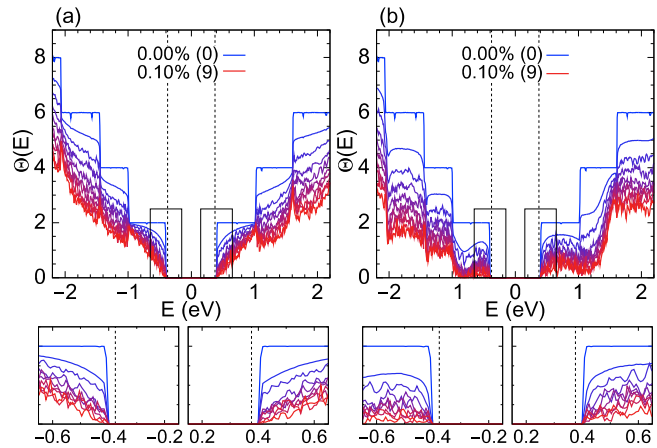

FIG. 3. (Color online) Transmission function of 100-nm CNTs with (a) vacancies and (b) SW defects. The bottom panels show the transmission of low-energy electrons, the blow-ups of the marked areas. The defect concentration varies 0.0% (blue) to 0.1% (red) ($N_{def} = 0$ to 9) with the equal interval.

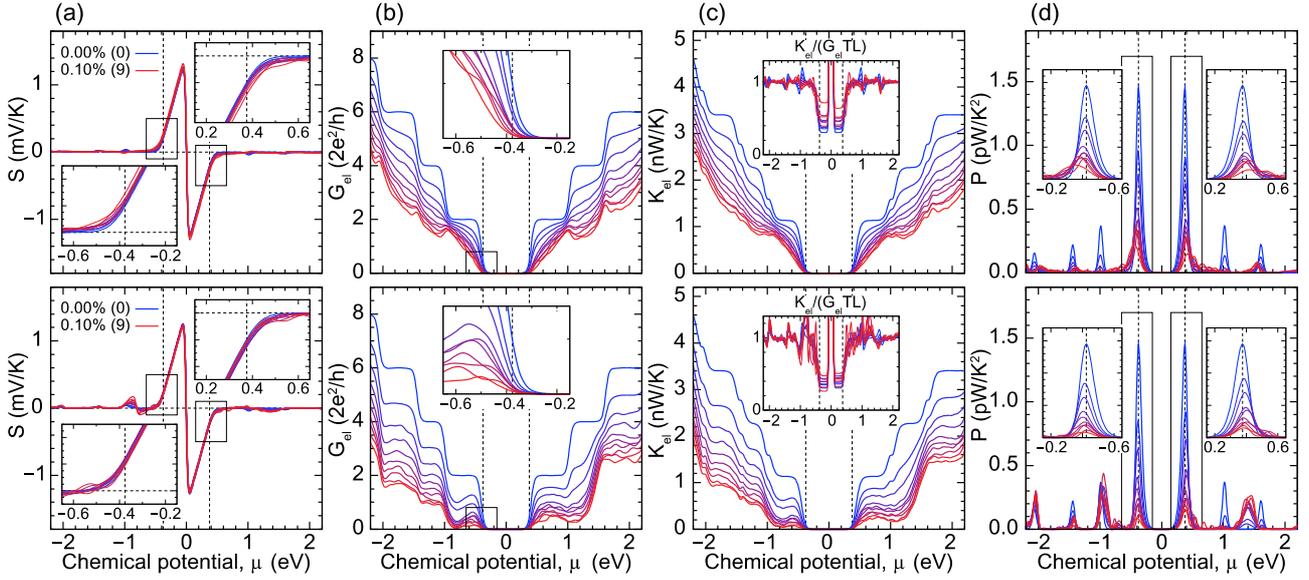

FIG. 4 (Color online) Thermoelectric properties of 100 nm-CNTs with vacancies (top) and SW defect (bottom): (a) Seebeck coefficient, (b) electron conductance, (c) electron thermal conductance, and (d) power factor. Color notification is the same as in Fig. 3. Dashed lines denote the peak chemical potentials for $P$, $\mu_{p/n,opt} = -/+ 0.38$ eV. Insets of (a), (b), and (d) show blow-ups of the marked region while those of (d) show the Wiedemann-Franz law.

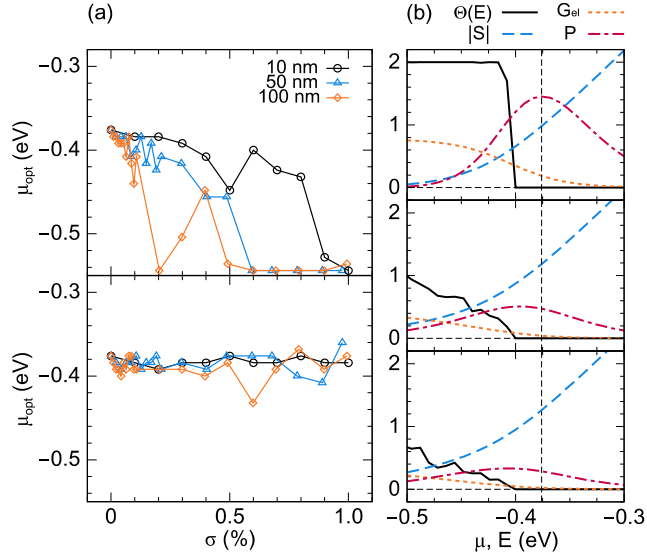

FIG. 5 (Color online) Peak chemical potential for $P$ shifts with increasing the defect concentration. (a) Fluctuation of the optimized chemical potential for $P$ of p-type CNTs with (top) vacancy and (bottom) SW defect. (b) Change in different electron transport properties for p-type CNTs with vacancies of $\sigma = 0.00$, 0.02, and 0.04% ($N_{def} = 0$, 3, and 6). The units for $|S|$, $G_{el}$, $P$, and $\Theta(E)$ are V/(5000K), S/5000, pW/K$^2$, and dimensionless, respectively. The competing behavior of $G_{el}$ and $S$ with the introduction of vacancies, the decrease in $G_{el}$ and increase in $S$, causes the shift of the peak chemical potential for $P$ toward high-doping level.

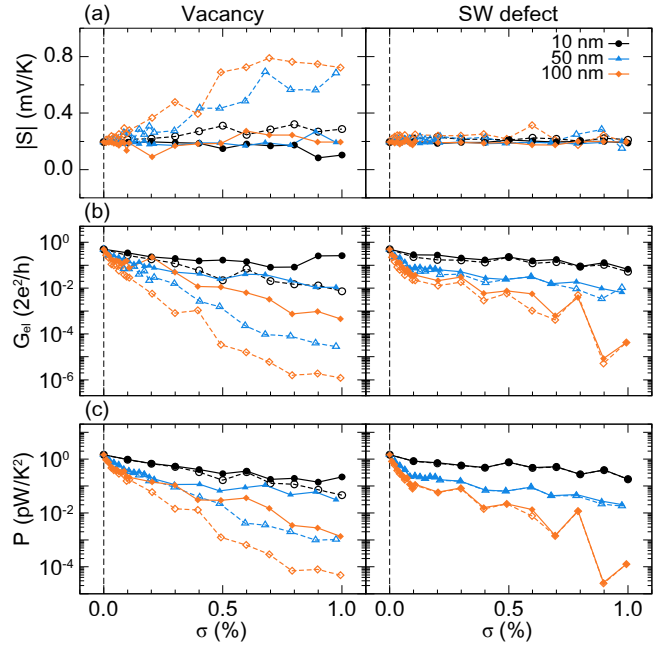

FIG. 6 (Color online) Thermoelectric properties of CNTs with vacancies (left) and SW defects (right): (a) $|S|$, (b) $G_{el}$, and (c) $P$. $P$ and $G_{el}$ are plotted on logarithmic scale. Solid and dashed lines show data at $\mu_{opt}$ and $\mu_0$, respectively. Orders of reduction of $G_{el}$ is a dominant factor of the reduction of $P$ due to defects.

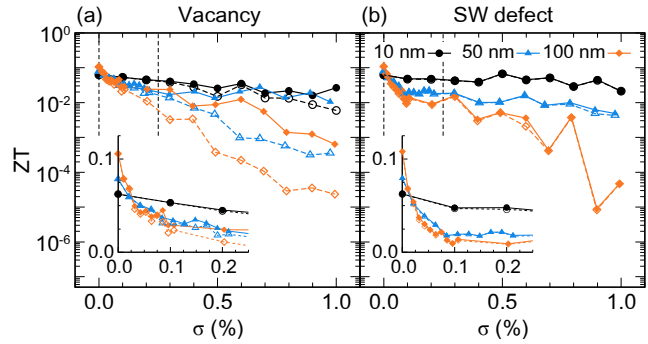

FIG. 7 (Color online) Figure of merit of CNTs with (a) vacancies and (b) SW defects. Insets show the data at low $\sigma$ (< 0.25%), denoted by dashed lines in the main figure, with linear scale. Solid and dashed lines show data at $\mu_{opt}$ and $\mu_0$, respectively, same as in Fig. 6.

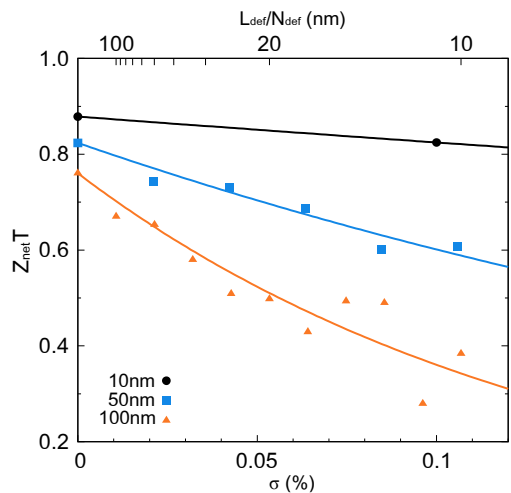

FIG. 8 (Color online) Variation of the figure of merit of CNT-based networks due to the introduction of vacancies.